\documentclass[twocolumn,showpacs,amsmath,amstex,amssymb,mathfonts,prl]{revtex4-1}
\usepackage{amsthm,amsfonts,graphicx,verbatim}

\usepackage{dcolumn}   
\usepackage{verbatim}
\usepackage{tipa}
\usepackage{wasysym}
\usepackage{esint}
\usepackage{color}

\usepackage{amsmath}
\usepackage{amssymb}
\usepackage{amsthm}
\usepackage{amsfonts}
\usepackage{listings}
\usepackage{enumerate}
\usepackage{latexsym}
\usepackage{subfigure}

\begin{document}

\title{Analytic Wavefunctions for Collective Modes in Fractional Quantum Hall Fluids}

\author{ Bo Yang$^1$}
\affiliation{$^1$ Department of Physics, Princeton University, Princeton, NJ 08544, USA}

\date{\today}
\begin{abstract}
We show model wavefunctions for neutral collective modes in fractional quantum Hall (FQH) states have simple analytic forms obtained from judicially reducing the powers of selected pairs in the ground state Jastrow factor. This scheme of ``pair excitations" works for the magneto-roton modes of single-component Abelian and non-Abeliean FQH states, as well as neutral fermion mode for the Moore-Read (MR) state. The analytic wavefunctions enable computations in the thermodynamic limit previous inaccessible to numerics, and the long wavelength limit of the neutral energy gap of the magneto-roton modes can be interpreted as fusion of charges in two-dimensional plasma picture, extending the plasma analogy to neutral excitations. A lattice diagrammatic method of representing these many-body wavefunctions and FQH elementary excitations is also presented.
\end{abstract}

\maketitle \vskip2pc

The fractional quantum Hall effect (FQHE)\cite{tsg} is one of the prime examples where strong interaction between electrons dictates the dynamics. A very fruitful approach for such a strongly correlated system is to look for model wavefunctions and model Hamiltonians, where physical interpretations are more transparent, that are adiabatically connected to experimentally accessible systems. For the ground states and charged excitations of FQHE at several filling factors, wavefunctions can be written down analytically\cite{l,m,r}. We can thus identify the properties of FQHE in the thermodynamic limit, and reinterpret the wavefunctions in analogy to two-dimensional plasmas\cite{l}, or as conformal blocks of conformal field theory (CFT)\cite{m,bonderson}. In general for these approaches, incompressibility of FQHE is always assumed, and the dynamics is not explicitly discussed.

Incompressibility of FQHE is defined by the neutral collective excitations. Such excitations are important for understanding which topological phases of FQHE can be stabilized, and also very relevant to recent development of fractional Chern insulator, where the issue of incompressibility and its mapping to FQHE\cite{ss,sm,wjs,roy,lb,wrb,ltq} are areas of active research. The first formal treatment of neutral excitations came from single mode approximation (SMA) for the magneto-roton\cite{gmp} mode, where good model wavefunctions of density wave excitations can be constructed numerically up to the momentum of roton minimum, beyond which SMA is no longer valid. Following experimental studies of the collective modes by several groups\cite{west1,foxon,cheng,west2}, more recently the model wavefunctions are constructed numerically both for the magneto-roton modes in the Read-Rezayi (RR) series, and the neutral fermion mode\cite{moller} in the Moore-Read (MR) state characteristic of its non-Abelian nature. These wavefunctions are good for the entire range of momentum. However in practice the long wavelength limit is not accessible due to limitation by the system size. It is now understood that the collective modes can be thought as excitons of composite fermions\cite{jain1,jain2,rodriguez}, and in the long wavelength limit it is a ``spin 2" quadrupole excitation; as momentum increases it relaxes into a dipole excitation beyond the roton minimum\cite{bernevig1,yzz}. Interestingly, even though the underlying phenomenological pictures of the collective modes can be different, it is found numerically that exactly the same set of model wavefunctions are found with different approaches, with very rich algebraic structures. This suggests a natural way of representing these model wavefunctions for the neutral excitations as well, with no need of explicit variational parameters.

In this Letter, we present analytic wavefunctions that are identical to those numerically generated in \cite{yzz,jain1,rodriguez,jain2}, and calculate the energy gap of the quadrupole excitations in the thermodynamic limit. We start by presenting the wavefunctions of the collective modes for fermionic Laughlin state at filling factor $\nu=1/m$ in the lowest Landau level (LLL), where $m$ is odd. On the sphere the ground state is the Laughlin wavefunction in total angular momentum $L=0$ sector, or the fermionic Jack polynomial  $J^{-m+1}_{1001001\cdots}$\cite{bernevig}. By stripping away the single particle normalization factor the holomorphic part of the wavefunction is the same on all 0-genus manifold. Thus we label the state with its total angular momentum on the sphere. Taking $z=\frac{1}{\sqrt{2}l_B}(x+iy), z^*=\frac{1}{\sqrt{2}l_B}(x-iy)$, where $l_B=\sqrt{\hbar/eB}$ is the magnetic length, the unique ground state is given by $\psi_l=\prod_{i<j}(z_i-z_j)^me^{-\frac{1}{2}\sum_iz_iz_i^*}$, with model Hamiltonian made of Haldane pseudopotentials\cite{haldane} $V=\sum_{i<j}V_{ij}$, with

\begin{eqnarray}\label{vij}
V_{ij}=\int \frac{d^2ql_B^2}{2\pi}\sum_{n=0}^{m-1}L_n(q^2l_B^2)e^{-\frac{1}{2}q^2l_B^2}e^{i\vec q\cdot (\vec R_i-\vec R_j)}
\end{eqnarray}

where $L_n(x)$ is the $n^{\text{th}}$ Laguerre polynomial and $\vec R_i$ is the guiding center coordinate of the $i^{\text{th}}$ particle. Physically, $V_{ij}$ is the short range interaction that projects into the two-body Hilbert space with their relative angular momentum smaller than $m$. The family of collective modes at different angular momentum sector (we omit the exponential part of the wavefunction, which is irrelavent in LLL) is as follows:

\footnotesize
\begin{widetext}
\begin{eqnarray}\label{collect}
&&\mathcal A[(z_1-z_2)^{m-2}\prod'_{i<j}(z_i-z_j)^m]\qquad L=2\nonumber\\
&&\mathcal A[(z_1-z_2)^{m-2}(z_1-z_3)^{m-1}\prod'_{i<j}(z_i-z_j)^m]\qquad L=3\nonumber\\
&&\mathcal A[(z_1-z_2)^{m-2}(z_1-z_3)^{m-1}(z_1-z_4)^{m-1}\prod'_{i<j}(z_i-z_j)^m]\qquad L=4\nonumber\\
\vdots
\end{eqnarray}
\end{widetext}
\normalsize

Here $\mathcal A$ indicates antisymmetrization over all particle indices, and $\prod'_{i<j}$ means products of only pairs $\{ij\}$ that do not appear in the prefactors to the left of it. Thus the $L=2$ state, which is the quadrupole excitation in the thermodynamic limit\cite{yzz}, is obtained from the ground state by reducing the power of one pair of particles (which we choose arbitrarily as particle $1$ and $2$) by \emph {two}, followed by antisymmetrizing over all particles. This scheme naturally forbids an $L=1$ state by pair excitation, since if we reduce the power of one pair of particles by \emph {one}, antisymmetrization kills the state. 

The $L=3$ state is generated by pairing particle $1$ with another particle and reducing their pair power by one. It is now clear how the modes in other momentum sectors are generated. Naturally for a total of $N_e$ particles, the family of collective modes ends at $L=N_e$, agreeing with the scheme in \cite{yzz}. Indeed all wavefunctions here satisfy the highest weight condition, and the states relax to the ground state far away from the excited pairs; these are exactly the conditions we used to numerically generate the unique model wavefunction in each momentum sector.

\begin{figure}
\includegraphics[width=3.5cm,height=1.5cm]{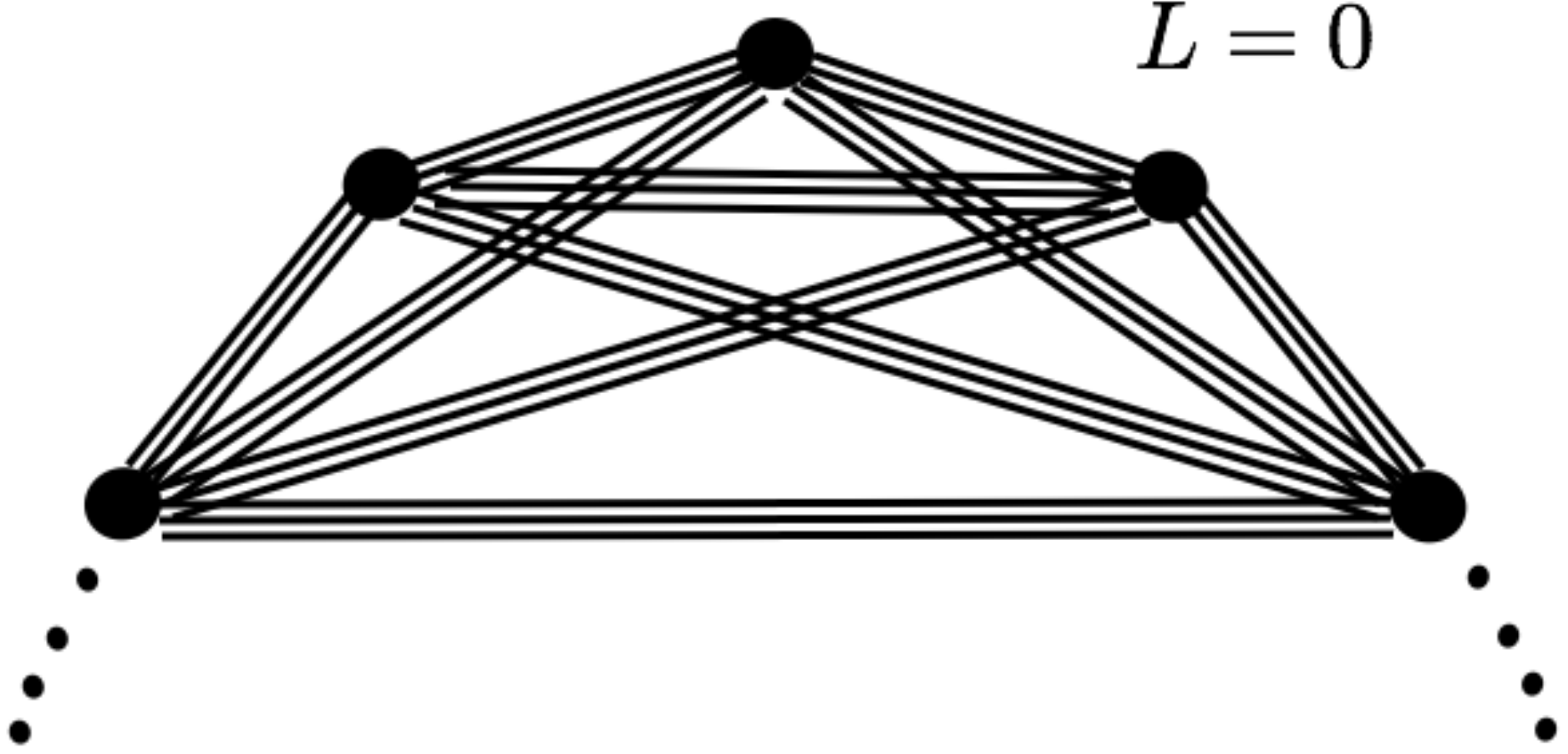}
\caption{For $N_e$ particles, the lattice can be viewed as an N-gon, with three bonds connecting every pair of vertices}
\label{fig:laughlin0}
\end{figure}

An intuitive way to visualize the family of collective modes is to map the particles onto a lattice, where each lattice site represents a particle. Since for FQHE we have a quantum fluid instead of a solid, every two lattice sites interact with each other. The number of bonds between each pair of lattice sites equal to the power of the pair of particles in the wavefunction. As an example we consider the simpliest Laughlin state at $\nu=1/3$, so for the ground state every two lattice sites are connected by three bonds, as shown in Fig.~\ref{fig:laughlin0}.

\begin{figure}
\includegraphics[width=5.5cm,height=3.5cm]{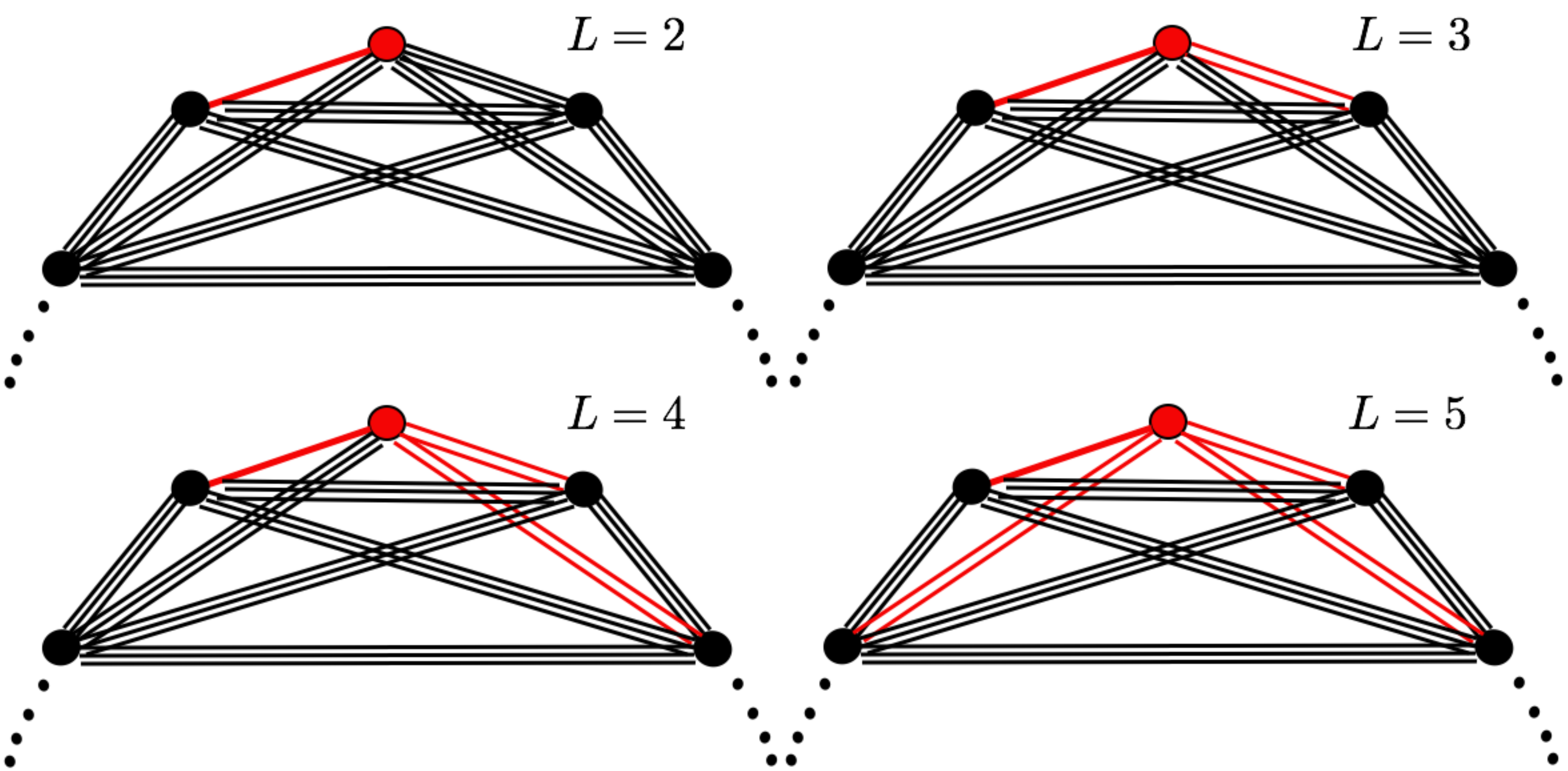}
\caption{(Color Online) Collective modes from $L=2$ to $L=5$, where the change of bonds are highlighted with red color.}
\label{fig:laughlin25}
\end{figure}

The collective modes are obtained by breaking the bonds between lattice sites, as shown in Fig.~\ref{fig:laughlin25}. We can view the entire family of collective modes as elementary excitations centered around a single red lattice site. Note the lattice pattern uniquely defines the many-body wavefunction, and different types of ``elementary excitations" can be identified with different patterns of bond-breaking around a single lattice site.

The same scheme applies to MR state. It is instructive to first see how the MR ground state is obtained. The Laughlin wavefunction at half filling is given by $J^{-2}_{1010101\cdots}(z_i)=\prod_{i<j}(z_i-z_j)^2$. For fermions this is not a valid state; instead the ground state was constructed by a pairing mechanism\cite{m}, which is also a Jack polynomial $J^{-3}_{1100110011\cdots}$. The pairing reduces the power of each pair of particles by \emph{one}. Explicitly we have for $2n$ particles

\begin{figure}
\includegraphics[width=5.5cm,height=1.5cm]{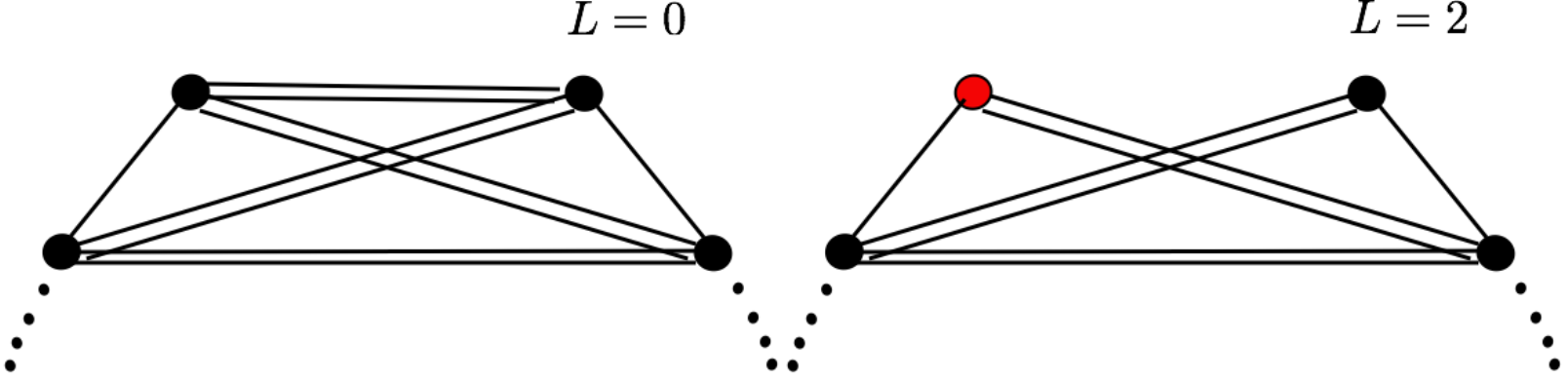}
\caption{(Color Online) The lattice configuration of the ground state $L=0$ and the first collective mode $L=2$. Consecutive collective modes can be obtained by breaking one of the \emph {double bonds} connecting the red lattice site to some other site.}
\label{fig:mr02}
\end{figure}

\footnotesize
\begin{eqnarray}\label{pairing}
&&\prod_{i<j}(z_i-z_j)^2\rightarrow\mathcal A[(z_1-z_2)(z_3-z_4)\cdots(z_{2n-1}-z_{2n})\prod'_{i<j}(z_i-z_j)^2]\nonumber\\
&&=\text{Pf}\left(\frac{1}{z_i-z_j}\right)\prod_{i<j}(z_i-z_j)^2
\end{eqnarray}
\normalsize

where the last line of Eq.(\ref{pairing}) is the familiar Pfaffian for the MR ground state. This suggests lattice representation of MR state and its magneto-roton mode with the same scheme, as shown in Fig.~\ref{fig:mr02}

We can also use the same scheme to generate the neutral fermion mode for the MR states with odd number of particles. In this case, starting from the Bosonic Laughlin wavefunction at half filling, every two particles form a pair except for one particle. Naturally the ``ground state" of the neutral fermion mode is given by

\small
\begin{eqnarray}\label{opairing}
&&\prod_{i<j}(z_i-z_j)^2\rightarrow\mathcal A[(z_1-z_2)(z_3-z_4)\cdots(z_{2n-1}-z_{2n})\nonumber\\
&&\prod_{k<2n+1}(z_{2n+1}-z_k)^2\prod'_{i<j}(z_i-z_j)^2]
\end{eqnarray}
\normalsize

\begin{figure}
\includegraphics[width=3.5cm,height=4.5cm]{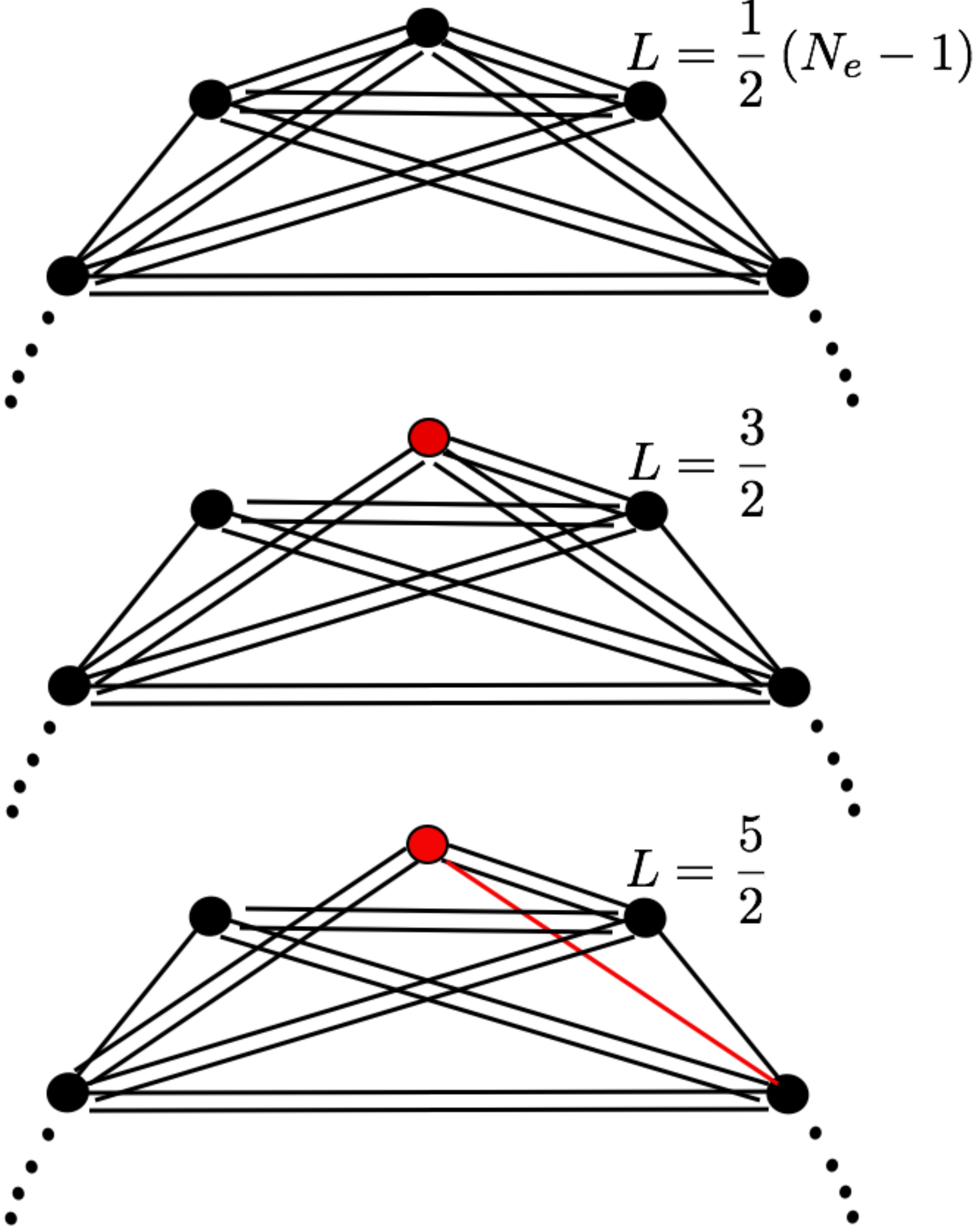}
\caption{(Color Online) The lattice configuration of the zero mode quasihole state $L=\frac{1}{2}\left(N_e-1\right)$ and the first two neutral fermion modes at $L=\frac{3}{2}\text{ and }L=\frac{5}{2}$. Consecutive collective modes can be obtained by breaking one of the \emph {double bonds} connecting the red lattice site to some other site.}
\label{fig:mrn02}
\end{figure}

For model three-body Hamiltonian, this is a zero-energy abelian quasihole state $J^{-3}_{1100110011\cdots 0011001}$ in the angular momentum sector $L=\frac{1}{2}(N_e-1)$. Its lattice presentation and that of the neutral fermion mode is given in Fig.~\ref{fig:mrn02}

To write down the analytic wavefunctions in a more formal way, we define $\mathcal P_{ij}=\frac{1}{z_i-z_j}$. Notice the Pfaffian for $2n$ particles can be written as $\text{Pf}\left(\frac{1}{z_i-z_j}\right)\sim\mathcal A[\mathcal P^{(2n)}]$, where $\mathcal P^{(2n)}=\mathcal P_{12}\mathcal P_{34}\cdots\mathcal P_{2n-1,2n}$. The magneto-roton mode for the Laughlin state is given by

\begin{eqnarray}\label{laughlinm}
\psi_l^{L=k+2}=\prod_{i<j}^{N_e}(z_i-z_j)^m\mathcal S[\mathcal P_{12}^2\mathcal P_{13}\cdots\mathcal P_{1,2+k}]
\end{eqnarray}

where $\mathcal S$ is the symmetrization over all particle indices. From the Bosonic Laughlin wavefunction at filling factor $1/2$ we can impose pairing to obtain

\begin{eqnarray}\label{mrg}
\psi_{\text{mr}}=\prod_{i<j}^{N_e}(z_i-z_j)^2\mathcal A[\mathcal P^{(2n)}]
\end{eqnarray}

For even number of electrons we have $N_e=2n$ and Eq.(\ref{mrg}) is the MR ground state. The magneto-roton modes are given by

\small
\begin{eqnarray}\label{mrm}
\psi_{\text{mr}}^{L=k+2}=\prod_{i<j}^{N_e}(z_i-z_j)^2\mathcal A[\mathcal P^{(2n)}\mathcal P_{13}^2\mathcal P_{15}\cdots \mathcal P_{1,3+2k}]
\end{eqnarray}
\normalsize

 For odd number of electrons we have $N_e=2n+1$ and Eq.(\ref{mrg}) is the MR quasihole state of Eq.(\ref{opairing}). The neutral fermion modes are given by

\footnotesize
\begin{eqnarray}\label{mrn}
\psi_{\text{mr}}^{L=\frac{3}{2}+k}=\prod_{i<j}^{N_e}(z_i-z_j)^2\mathcal A[\mathcal P^{(2n)}\mathcal P_{N_e,1}^2\mathcal P_{N_e,3}\cdots\mathcal P_{N_e,1+2k}]
\end{eqnarray}
\normalsize

The analytic wavefunction is useful in calculating the magneto-roton mode energy gap in the long wavelength limit. For the Laughlin state, the energy gap is given by

\begin{eqnarray}\label{energy}
\epsilon_{q\rightarrow 0}=\lim_{N_e\rightarrow\infty}\frac{\langle\psi^{L=2}_l|V|\psi^{L=2}_l\rangle}{\langle\psi^{L=2}_l|\psi^{L=2}_l\rangle}
\end{eqnarray}

We already know from \cite{yzz} that in $L=2$ and $L=3$ sector, SMA is exact for the magneto-roton mode model wavefunction. Defining the guiding center ladder operators as $b^\dagger_i=z_i,b_i=\partial_{z_i}$, we have

\begin{eqnarray}\label{sma}
\psi^{L=2}_l&=&\frac{1}{2m(m-1)}\sum_i(b_i)^2\psi_l
\end{eqnarray}

In the thermodynamic limit, the normalization constant of the above two modes are thus related to the long wavelength expansion of the ground state guiding center structure factor:

\begin{eqnarray}\label{sq}
S_q&=&\frac{1}{N_\phi}\left(\langle\delta\bar\rho_q\delta\bar\rho_{-q}\rangle_0-\langle\delta\bar\rho_q\rangle_0\langle\delta\bar\rho_{-q}\rangle_0\right)\nonumber\\
&=&-\frac{\bar s}{4m}(g^{ab}q_aq_b)^2+O(q^6)
\end{eqnarray}

where $\bar s=\frac{1-m}{2}$ is the guiding center spin\cite{gmp,haldane}, and $g^{ab}$ is the guiding center metric\cite{yzrh}. We thus have 

\begin{eqnarray}\label{denominator}
\langle\psi^{L=2}_l|\psi^{L=2}_l\rangle=-\frac{\bar sN_e}{2m^2(m-1)^2}
\end{eqnarray}

The numerator of Eq.(\ref{energy}) can be calculated with plasma analogy. Note in Eq.(\ref{collect}), before antisymmetrization the term only has one pair of particles with relative angular momentum smaller than $m$. We thus have

\small
\begin{eqnarray}\label{onepair}
\langle\psi^{L=2}_l|V|\psi^{L=2}_l\rangle&=&\frac{N_e(N_e-1)}{2\mathcal N^2}\langle\bar\psi_l|\mathcal P_{12}^2V_{12}\mathcal P^2_{12}|\bar\psi_l\rangle
\end{eqnarray}
\normalsize

where $\mathcal N$ is the normalization constant of the Laughlin state. We note that $V_{12}$ projects out states with relative angular momentum $(z_1-z_2)^{m-2}$, which can be integrated over. The numerator is thus equivalent to evaluating the norm of the following wavefunction:

\small
\begin{eqnarray}\label{numerator}
\bar\psi=\prod_{i=2}^{N_e-1}\left(\frac{1}{\sqrt{2}}z_1-z_i\right)^{2m}\prod_{1<i<j<N_e-1}(z_i-z_j)^m
\end{eqnarray}
\normalsize

which can be evaluated as the free energy of two-dimensional one-component plasma (OCP) on a disk with radius $R^2=\frac{mN_e}{2}$ and elementary charge $e=2\sqrt{\pi mk_BT}$, where particle $1$ interacts with the rest of the particles with charge $2e$. We thus obtain

\begin{eqnarray}\label{energyr}
\epsilon_{q\rightarrow 0}=-\frac{2^mm(m-1)^2}{\pi\bar s}e^{-\frac{\mathcal F_2-\mathcal F}{k_BT}}
\end{eqnarray}

Both $\mathcal F_2\text{ and }\mathcal F$ are free energies of OCP in the thermodynamic limit ($N_e\rightarrow\infty$), where $\mathcal F$ is for $N_e$ particles, each with charge $e$ with logarithmic two-body interactions together with a neutralizing background of radius $R$; for $\mathcal F_2$, we have the same neutralizing background with $N_e-2$ particles of charge $e$, and exactly one particle with charge $2e$. Thus $\mathcal F_2-\mathcal F$ is the free energy cost of fusing two particles of charge $e$ to create a particle of charge $2e$, which is an $O(1)$ effect.

Similar calculations can be carried out for the magneto-roton mode in the MR state. Analogous to Eq.(\ref{sma}) we have $\psi^{L=2}_{\text{mr}}=\frac{1}{24}\sum_ib_i^2\psi_{mr}$,  and in the long wavelength limit we have

\begin{eqnarray}\label{energymr}
\epsilon^{\text{mr}}_{q\rightarrow 0}=-\frac{24}{\pi\bar s_{\text{mr}}}e^{-\frac{\mathcal F_3-\mathcal F_{\text{II}}}{k_BT}}
\end{eqnarray}

where  $\bar s_{\text{mr}}=-2$ is the guiding center spin for the MR state, and $\mathcal F_{\text{II}}$ is the standard two-component plasma free energy for the MR ground state\cite{bonderson}. The charge for interaction between the two components is given by $Q_1=\pm \sqrt{3k_BT}$, while the charge for interaction between one component and the neutralizing background is given by $Q_2=2\sqrt{k_BT}$. $\mathcal F_3-\mathcal F_{\text{II}}$ is the free energy cost of fusing \emph {three particles} for each component to create one particle with charge $3Q_2$ but with the same $\pm Q_1$. 

The evaluation of the long wavelength gap of the neutral fermion mode is less transparent. The difficulty lies with evaluating the normalization constant of $\psi^{L=\frac{3}{2}}_{\text mr}$. There is no SMA for the neutral fermion mode, and it is not known if in the thermodynamic limit the gap should be inversely proportional to the guiding center spin. On the other hand $\langle\psi^{L=\frac{3}{2}}_{\text{mr}}|V_{\text{3bdy}}|\psi^{L=\frac{3}{2}}_{\text{mr}}\rangle$ can be mapped to 2-component plasma as well, and we obtain

\begin{eqnarray}\label{energymrnf}
\bar\epsilon^{mr}_{q\rightarrow 0}\sim e^{-\frac{\bar{\mathcal F}_3-\bar{\mathcal F}_{\text{II}}}{k_BT}}
\end{eqnarray}

Here $\bar{\mathcal F}_{\text{II}}$ is the free energy of the 2-component plasma similar to that of $\mathcal F_{\text{II}}$ with only one difference: there is exactly one \emph {more} particle carrying charge $Q_2$ that interacts with the neutralizing background, and its $Q_1$ charge is zero. This is how an unpaired fermion in the MR state is interpreted in the plasma analogy. Furthermore, $\bar{\mathcal F}_3-\bar{\mathcal F}_{\text{II}}$ is the energy cost of fusing the unpaired fermion with one pair of two other fermions, creating a particle with charge $Q_2=6\sqrt{k_BT}$ but again with \emph zero $Q_1$. 

In conclusion, analytic wavefunctions for both the magneto-roton modes and the neutral fermion modes are presented. The energy gap of the quadrupole excitation in the thermodynamic limit can be related to the free energy cost of the fusion of charges in the plasma energy, and is inversely proportional to the guiding center spin which characterizes its topological order. This is the first time that the plasma analogy is extended to neutral excitations of FQHE, and the analogy not only applies to the wavefunctions, but also to the dynamics as well.  The lattice diagrams we presented uniquely defines the many-body wavefunctions; one would also conjecture the diagrams are useful in determining the many-body wavefunctions of multi-roton excitations. Since the collective mode in the long wavelength limit is buried in the multi-roton continuum, it is important to calculate the decay rate of the collective mode. Numerical calculation has been performed to show that even in the continuum the decay rate of the collective mode is very small. This opens up the possibility of experimental detection of these modes. A more detailed analysis of the decay rate of collective modes will be presented elsewhere\cite{yz}.

{\sl Acknowledgements}. I would like to thank F.D.M Haldane for many useful discussions, and Zlatko Papic for help in doing numerical evaluations. I also thank Chris Laumann for useful discussions and pointing me to the work of \cite{bonderson}. This work was supported by DOE grant DE-SC$0002140$ and NSS Scholarship by ASTAR.

\end{document}